\documentclass[a4paper,final]{aipproc}

\usepackage{epsfig,graphicx,epsf}
\usepackage{amssymb,amsfonts,amsmath}

\layoutstyle{8x11single}

\newcommand{\be}{\begin{equation}}
\newcommand{\ee}{\end{equation}}
\newcommand{\no}{\nonumber}

\newcommand{\ba}{\begin{eqnarray}}
\newcommand{\ea}{\end{eqnarray}}
\newcommand{\bal}{\begin{aligned}}
\newcommand{\eal}{\end{aligned}}

\begin{document}

\title{Pulsar Timing Arrays and the cosmological constant}


\author{Dom\`enec Espriu}{
 address={Departament d'Estructura i Constituents de la Mat\`eria and 
Institut de Ci\`encies del Cosmos (ICCUB), Universitat de Barcelona,
Mart\'\i ~i Franqu\`es, 1, 08028 Barcelona, Spain.}
}

\begin{abstract}
In this talk I review how a non-zero cosmological constant $\Lambda$ affects
the propagation of gravitational waves and their detection in pulsar timing arrays (PTA). 
If $\Lambda\neq 0$ it turns out that waves are anharmonic in cosmological Friedmann-Robertson-
Walker coordinates and although the amount of anharmonicity is very small it leads to 
potentially measurable effects. The timing residuals induced by gravitational waves in PTA 
would show a peculiar angular dependence with a marked enhancement around a particular 
value of the angle subtended by the source and the pulsars. This angle depends mainly on the actual 
value of the cosmological constant and the distance to the source. Preliminary estimates
indicate that the enhancement can be rather notorious for supermassive black hole mergers 
and in fact it could facilitate the first direct detection of gravitational waves while 
at the same time representing a `local' measurement of $\Lambda$.
\end{abstract}

\maketitle

\vspace{-10cm}\begin{flushright}{ICCUB-14-032}
\end{flushright}\vspace{9cm}


\section{Introduction}
There is a fair amount of evidence suggesting that the space time where we live is globally de Sitter with
a value for the cosmological constant estimated \cite{concordance} to be around
$\Lambda \simeq 10^{-52}$ m$^{-2}$. This has obvious effects on cosmology -- at very large scales. These effects
are of no concern to us here.

Instead we would like to help answer the question: does $\Lambda$ have a `local' influence?, where 
local means at moderate values of the redshift $z$. In short, can the cosmological constant be measured `locally'?
This is an important question because it may settle the issue as to whether $\Lambda$ is a truly 
fundamental property of space-time, a basic constant of nature present at {\em all} scales,
rather than a way of providing some effective description relevant only at cosmological distances.

A lot of work has been devoted to finding traces of the existence of the cosmological constant at sub-cosmological
scales such as in cluster of galaxies \cite{previous}, so far without a clear conclusion or very relevant bounds. 
Several works study the effect of $\Lambda$ on the gravitational bending of ligh \cite{KP,Sereno,RindlerIshak} or the
Shapiro effect \cite{bernabeunick}. Some authors have even advocated more exotic effects such as
providing an explanation the  well known Pioneer 
anomaly\cite{pioneer} or the apparent increse of the astonomical unit with time\cite{AU}.

Discussions in the literature regarding the above points tend to be confusing. Effects range from surprisingly large to 
zero. Not surprisingly the source of most discrepancies is the meaning of the different coordinate systems and their
physical realization. 

Here we propose to try to find `local' effects of $\Lambda$ by studying the propagation of
gravitational waves (GW) in an space
endowed with a cosmological constant. This may seem hopeless at first as there is no direct 
detection of a GW yet, let alone possible modifications due to $\Lambda\neq 0$. However because the nature of 
the cosmological constant is quite unclear it is plausible that it should be attributed to 
the gravitational interaction itself ---a fundamental property of space-time--- and accordingly placed 
naturally on the l.h.s. of Einstein equations.
Then it seems quite natural to investigate how the fundamental excitations of gravity are modified by $\Lambda$.
This issue has been studied in \cite{BEP}.

Has $\Lambda$ possible observational consequences on GW in spite of its currently preferred 
very small value? We will see that the answer to this question is somewhat surprising. What follows is
an extended version of the results found by us and described in \cite{EP}.

For a previou attempt to find effects of a non-zero cosmological constant using GW see \cite{nafjetzer}.

\section{Linearization of the field equations in the presence of $\Lambda$}
Keeping control of the different orders in $\Lambda$ will be essential to our discussion. This will
become obvious in the subsequent.

Let us start by linearizing Einstein equations since after all GW are solutions of the linearized
equations:
\be
R_{\mu\nu}-\frac{1}{2} g_{\mu\nu}R+\Lambda g_{\mu\nu}=-\kappa T_{\mu\nu} 
\ee
In the linearized approximation we assume $g_{\mu\nu}=\eta_{\mu\nu}+h_{\mu\nu}, \, |h_{\mu\nu}| \ll 1$ and then
\be
R_{\mu\nu}=\frac{1}{2}\left(\Box h_{\mu\nu}+h_{,\mu\nu}-h_{\mu ,\nu\lambda}^{\lambda}-
h_{\nu ,\mu\lambda}^{\lambda}\right).
\ee
A gauge choice is mandatory to solve these equations. A common choice is to select the Lorenz gauge
\be
\partial_{\mu}h_{\,\nu}^{\mu}=\frac{1}{2}\partial_{\nu}h \leftrightarrow
\partial_{\mu}\tilde h_{\,\nu}^{\mu}=0,
\ee
where
\be
\tilde h _{\mu\nu} = h_{\mu\nu}-\frac12 \eta_{\mu\nu} h 
\ee
The field equation in Lorenz gauge are
\be\label{einsteinequation}
\Box \left(h_{\mu\nu}-\frac{1}{2}\eta_{\mu\nu}h\right) + 2\Lambda h_{\mu\nu}= -2\Lambda\eta_{\mu\nu}.
\ee
Whether the term of order $\mathcal{O}(h\Lambda)$ needs to be considered or not depends on the relative magnitude
of $\Box h$ and $\Lambda$.
If the $\Lambda h_{\mu\nu}$ term on the l.h.s. is omitted (and only in this case)
there is a residual gauge freedom within the Lorenz gauge.
\be
x^{\mu}\rightarrow x'^{\mu}=x^{\mu}+\xi^{\mu}
\ee
as long as $\xi^{\mu}$ is an harmonic function, $\Box \xi^{\mu}=0$.
All this is of course well known and standard texbook material. If we set $\Lambda$ to zero 
we get the standard treatment of gravitational waves ---in Minkowski space time. It is clear
that $\Lambda\neq 0$ will modify the for of the solution by terms of order $\Lambda$. But we also 
need to know what choice of coordinates the linearization and the Lorenz condition implies, if any. 
This is not a totally trivial issue in de Sitter.

\section{`Good' and `bad' coordinate systems in de Sitter}
Before giving the solution to the field equation in the Lorenz gauge (the one where GW are usually treated) 
let us discuss several possible coordinate choices in de Sitter space-time. Convenient references
are given in \cite{coordinates}. See also \cite{bernabeuold}.

A {\em static} metric:  Schwarzschild-de Sitter (SdS)
\be
d{ s}^2=\left[1-\frac{\Lambda}{3}{\hat r}^2\right]d{\hat t}^2-
\left[1-\frac{\Lambda}{3}{\hat r}^2\right]^{-1}{\hat r}^2+{\hat r}^2d\Omega^2
\ee
This metric has not quite spherical symmetry. 

A {\em position independent} metric: Friedmann-Robertson-Walker (FRW)
\be
ds^2= dT^2 -\exp(2\sqrt{\frac{\Lambda}{3}}T) d\vec X^2 
\ee
This metric incorporates the physical principles of cosmological homogeneity and isotropy.
The coordinates $X^i$ are comoving coordinates
anchored in space that expand with the universe. These are the coordinates where our
world appears homogeneous and isotropic. 

None of the previous metrics obey the Lorenz gauge condition. Of course there is no reason
why they should because they are not solutions of Einstein equations linearized around
Minkowski space-time, so in order to match with the discussion
in the previous section let us proceed to lineariarize the two metrics. 
For the SdS metric this is quite easy 
as the  SdS metric is expandable  in integer powers of $\Lambda$
\be
d{s}^2\simeq \left[1-\frac{\Lambda}{3}{\hat r}^2\right]d{\hat t}^2-
\left[1+\frac{\Lambda}{3}{\hat r}^2\right]{\hat r}^2+{\hat r}^2d\Omega^2.
\ee
It should therefore obey a linearized version of Einstein equations, although as can be easily
verified not in the Lorenz gauge.

On the contrary it is clear that the FRW coordinate choice cannot be linearized in $\Lambda$ because it 
contains odd powers of $\sqrt\Lambda$
\be
ds^2= dT^2 -\exp(2\sqrt{\frac{\Lambda}{3}}T) d\vec X^2 \
\ee
and therefore it is impossible that it can fulfill
any linearized Einstein equation, even if $t<<1/\sqrt{\Lambda}$
as it is not expandable in integer powers of $\Lambda$.

One can work out the exact transformation between the two coordinate systems. We shall reserve capital letters
for FRW coordinates and the hatted lower case ones for SdS (the reason for the hat will be evident below)  
\be
\hat{r}=e^{T\sqrt{\Lambda /3}}R
\ee
\be
\hat{t}=\sqrt{\frac{3}{\Lambda}}\log\left(\frac{\sqrt{3}}{\sqrt{3-\Lambda e^{2T\sqrt{\Lambda /3}}R^2}}\right)+T
\ee
where $T$ and $R$ are the cosmological time and comoving coordinates whose physical realization
is clear. 
This transformation is valid inside the cosmological horizon, i.e. $R<\frac{1}{\sqrt{\Lambda}}$.

\section{Linearized background and linearized GW solutions}
Let us now return to Eq. \ref{einsteinequation}. We shall work in
the linearized approximation both for the background modification (with respect to flat Minkowski space-time)
$h_{\mu\nu}^\Lambda$ and for gravitational wave
perturbations $h_{\mu\nu}^W$. We follow here the discussions in \cite{BEP,bernabeuold}.
The total metric will be written as
\be
g_{\mu\nu}= \eta_{\mu\nu}+ h_{\mu\nu}^\Lambda + h_{\mu\nu}^W,\qquad |h_{\mu\nu}^{\Lambda,W}|\ll 1
\ee

Let us focus on the background. In the Lorenz gauge and neglecting $\Lambda h_{\mu\nu}^\Lambda$
\be
\Box \tilde{h}_{\mu\nu}=-2\Lambda\eta_{\mu\nu},\qquad
\partial_{\mu}\tilde{h}_{\nu}^{\mu}=0, 
\ee
This has as a particular solution
\be
\tilde h_{\mu\nu}=-\frac{\Lambda}{18}\left(4x_{\mu}x_{\nu}-\eta_{\mu\nu}x^2\right)\,\Rightarrow\,
h_{\mu\nu}=\frac{\Lambda}{9}\left(x_{\mu}x_{\nu}+2\eta_{\mu\nu}x^2\right)
\ee
The general solution is the former plus any solution of $\Box \tilde{h}_{\mu\nu}$ but we  
call the latter `waves' rather than `background'. Indeed 
the equation for GW is particularly simple in Lorenz coordinates if the $\Lambda h_{\mu\nu}^W$ term is neglected.
\be
\Box \tilde h_{\mu\nu}^W=0 
\ee
i.e. it is truly a wave equation.

Now we have to answer the following question:
What is the physical realization of the coordinates $x,t$ where we just solved Einstein equations in the
Lorenz gauge?

Let us take advantage of the residual gauge invariance 
that exists if the term $\Lambda h_{\mu\nu}$ is neglected. Then
the solution
\be
\tilde h_{\mu\nu}=-\frac{\Lambda}{18}\left(4x_{\mu}x_{\nu}-\eta_{\mu\nu}x^2\right) 
\ee
or
\be
h_{\mu\nu}=\frac{\Lambda}{9}\left(x_{\mu}x_{\nu}+2\eta_{\mu\nu}x^2\right)
\ee
can be transformed into a static metric --- still in Lorenz gauge.

The following change of coordinates
{\small
\be
x=x'+\frac{\Lambda}{9}\left(-t'^2-\frac{x'^2}{2}+\frac{(y'^2+z'^2)}{4}\right)x'
\ee
\be
y=y'+\frac{\Lambda}{9}\left(-t'^2-\frac{y'^2}{2}+\frac{(x'^2+z'^2)}{4}\right)y'
\ee
\be
z=z'+\frac{\Lambda}{9}\left(-t'^2-\frac{z'^2}{2}+\frac{(x'^2+y'^2)}{4}\right)z'
\ee
\be
t=t'-\frac{\Lambda}{18}(t'^2+r'^2)t' \no
\ee
}
transforms the metric into a static solution at order $\Lambda$
\be
ds^2=\left[1-\frac{\Lambda}{3}r'^2\right]dt'^2-\left[1-\frac{\Lambda}{6}(r'^2+3x_{i}'^2)\right]dx_{i}'^2.
\ee
This metric has a $Z_3$ symmetry only. We are still in the Lorenz gauge.

Under the following additional change
\be
x'=x''+\frac{\Lambda}{12}x''^3,\quad
y'=y''+\frac{\Lambda}{12}y''^3,\quad
z'=z''+\frac{\Lambda}{12}z''^3, 
\ee
\be
t'=t'' 
\ee
the metric becomes
\be
ds^2=\left[1-\frac{\Lambda}{3}r''^2\right]dt''^2-\left[1-\frac{\Lambda}{6}r''^2\right](dr''^2+r''^2d\Omega^2),
\ee
which is {\em not} in the Lorenz gauge anymore. Yet another change
\be
r''=\hat{r}+\frac{\Lambda}{12}\hat{r}^3\qquad
t''=\hat{t} \no
\ee
leads to
\be
d{s}^2=\left[1-\frac{\Lambda}{3}\hat{r}^2\right]d\hat{t}^2-\left[1+\frac{\Lambda}{3}\hat{r}^2\right]d\hat{r}^2+
\hat{r}^2d\Omega^2.
\ee
This is the linearized Schwarzschild-de Sitter metric.

Now we know in which coordinates we were when we solved the linearized Einstein equation 
with a cosmological constant in Lorenz gauge. A series of elementary coordinate transformations 
brought our solution to a linearized version of the 
Schwarzschild-de Sitter
metric (expanded to first order in $\Lambda$). Thanks to Birkhoff's theorem\cite{birkhoff}, we know 
that this metric is unique.

The SdS coordinates are useful for problems with spherical symmetry, such as objects falling onto each other.
They do not admit a Newtonian limit if $\Lambda\neq 0$, i.e.
\be
ds^2 \neq  (1+ 2\Phi)dt^2 - (1- 2\Phi) dx^2 
\ee
 because in addition to the scalar potential $\Phi$ (which itself has 
a correction of $O(\Lambda)$) there is a {\it tensor} potential 
$\tau_{ij} \sim \Lambda$. 
However it is still true that if a massive source of mass $M$ is introduced via
\be
T_{00} = M \delta(\hat r)
\ee
the equation of motion for a non-relativistic body are governed by $\Phi$ alone
\be
\ddot{\hat{r}}= - \frac{GM}{\hat r} + O(\Lambda) 
\ee
Solutions are periodic in coordinates $\hat r, \hat t$ (up to $O(\Lambda)$) and
the wave equation will lead to harmonic GW in {\it these coordinates} up to corrections of $O(\Lambda)$.

However, the coordinates where the metric is Schwarzschild-de Sitter 
centered in a remote black hole
are not `useful' for cosmology simply because we do not perform measurements here
using these.
But we know how to
go from these SdS coordinated to the ones (FRW) where cosmological measuments are made.
Of course none of these subtleties occur if $\Lambda=0$.

\section{Gravitational waves in cosmological coordinates}
Let us go back to Lorenz gauge. Recall that we want  $h_{\mu\nu}=h_{\mu\nu}^\Lambda + h_{\mu\nu}^W$ and that
$h_{\mu\nu}^W$ should fulfill, if the term $\Lambda h^W$ is neglected,
\be
\tilde h^{W\mu}_{\,\mu} = 0, \quad \partial_\mu\tilde h^{W\mu}_{\,\nu} =0, \quad
\Box \tilde h_{\mu\nu}^W = 0 
\ee
while
$h_{\mu\nu}^\Lambda$ should fulfill if $\Lambda h^\Lambda$ is neglected
\be
\tilde h^{\Lambda\mu}_{\,\mu} = 0, \quad \partial_\mu\tilde h^{\Lambda \mu}_{\,\nu} =0, \quad
\Box \tilde h_{\mu\nu}^\Lambda
= -2\Lambda\eta_{\mu\nu}.
\ee
The general solution at lowest order (written for $h_{\mu\nu}$) will be
\be
h_{\mu\nu}=h_{\mu\nu}^{\Lambda}+h_{\mu\nu}^{W}=
\frac{\Lambda}{9}\left(x_{\mu}x_{\nu}+2\eta_{\mu\nu}x^2\right)+E_{\mu\nu}^{W}\cos{kx}+D_{\mu\nu}^{W}\sin{kx}
\ee
where $E_{\mu\nu}$ and $D_{\mu\nu}$ are polarization tensors having vanishing traces
$E^{W}=D^{W}=0$ and obeying the condition $k_{\mu}E^{\mu W}_{\nu}=k_{\mu}D^{\mu W}_{\nu}=0$ with $k^2=0$.

It is possible to derive the full solution including $\Lambda h_{\mu\nu}$ terms but we shall not 
consider it here as the modifications are unrealistically small to be seen, but they have some
interesting aspects nevertheless. The interested reader can see \cite{BEP} for details.

This coordinate system is easily related to SdS coordinates. These coordinates are well suited to describe
problems with spherical symmetry ($\hat r =0$ is a `special' point) such as the solar system or 
collapse on to a black hole but these coordinates are not the ones
where we observe the (expanding) universe.
When the spherical symmetry is lost (away from the source), we have to match to suitable coordinates, 
physical to the observer.

We have to transform now the solutions found in SdS-like coordinates to FRW coordinates.
Let us show here for simplicity just how the lowest order solution (i.e. the one
obtained neglecting $\Lambda h^W_{\mu\nu}$ terms) looks once transformed. A plane wave 
propagating in the $\hat z$ direction transforms into  
\be\label{waveleading}
\bal 
h_{\mu\nu}^{W_{FRW}}=
&\begin{pmatrix}
  0 &  0 & 0 & 0  \\
  0 & E_{11}\left(1+2\sqrt{\frac{\Lambda}{3}}T\right) & E_{12}\left(1+2\sqrt{\frac{\Lambda}{3}}T\right) & 0 \\
 0  & E_{12}\left(1+2\sqrt{\frac{\Lambda}{3}}T\right) & -E_{11}\left(1+2\sqrt{\frac{\Lambda}{3}}T\right) &0 \\
    0   &0 &0 & 0 \\
    \end{pmatrix} \times \\
    &\cos{\left(w(T-Z)+w\sqrt{\frac{\Lambda}{3}}\left(\frac{Z^2}{2}-T Z\right)+\mathcal{O}(\Lambda)\right)}
+\mathcal{O}(\Lambda)\\
&+\begin{pmatrix}
  0 &  0 &0 & 0  \\
 0& D_{11}\left(1+2\sqrt{\frac{\Lambda}{3}}T\right) & D_{12}\left(1+2\sqrt{\frac{\Lambda}{3}}T\right) & 0 \\
0  & D_{12}\left(1+2\sqrt{\frac{\Lambda}{3}}T\right) & -D_{11}\left(1+2\sqrt{\frac{\Lambda}{3}}T\right) &0 \\
    0   &0 &0 & 0 \\
    \end{pmatrix} \times\\
    &\sin{\left(w(T-Z)+w\sqrt{\frac{\Lambda}{3}}\left(\frac{Z^2}{2}-T Z\right)+\mathcal{O}(\Lambda)\right)}
    +\mathcal{O}(\Lambda)
\eal
\ee
The maxima of the wave will be reached when
\be
w(T-Z)+w\sqrt{\frac{\Lambda}{3}}\left(\frac{Z^2}{2}-T Z\right)=n\pi 
\ee
\be
Z_{\text{max}}(n,T)\simeq T-\frac{n\pi}{w}-\frac{T^2}{2}\sqrt{\frac{\Lambda}{3}}+\frac{n^2\pi^2}{2w^2}\sqrt{\frac{\Lambda}{3}}
\ee
The phase velocity of the wave is
\be
v_{p}(T)\equiv\frac{dZ_{\text{max}}}{dT}=1-T\sqrt{\frac{\Lambda}{3}}+\mathcal{O}(\Lambda)
\ee
In comoving coordinates the phase velocity is smaller than 1. This does not mean that the waves slow down.
We can calculate the velocity in `ruler' distance.
\be
-dl^2=-\left(1+2T\sqrt{\frac{\Lambda}{3}}\right)dZ^2,\quad
\frac{dl}{dT}=\frac{d}{dT}\left[\left(1+T\sqrt{\frac{\Lambda}{3}}\right)dZ_{\text{max}}\right]\simeq 1
\ee
Notice that the modifications due to the cosmological constant are not of order $\Lambda$ as a naive consideration
of the linearized field equation (obeyed by the GW) would lead us to believe. As we discussed in much detail, this 
linearized equation in the Lorenz gauge is in an essential way related to SdS coordinates. The effect of the
coordinate change to the cosmological FRW coordinates is actually an effect of order $\sqrt\Lambda$ and therefore
the magnitude of the change can be very different. 

Several physical magnitudes appear in the modified expression
for the waves; in particular $\Lambda$. Its effects are shown in the accompanying figure and
the most relevant question is: are these changes detectable?
\begin{figure}[htbp]
\centering
\includegraphics[scale=0.55]{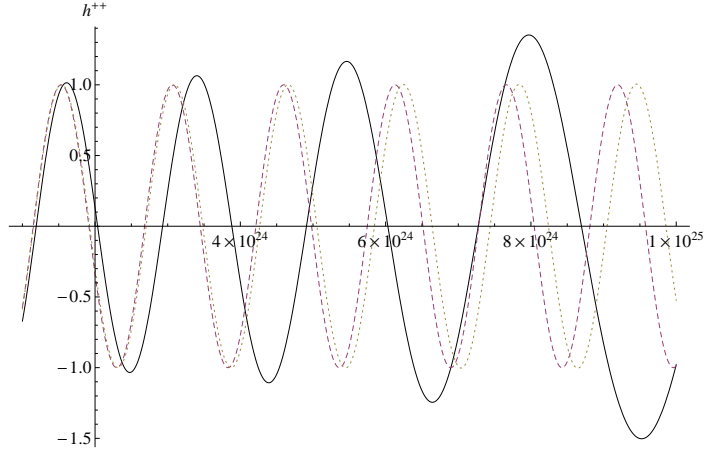}
\caption{Dependence of the amplitude and wave-length on $Z$ (expressed in meters) for a constant value of $T$ 
and for different values of $\Lambda$. Dashed line: $\Lambda=0$, dotted line: $\Lambda=10^{-52}m^{-2}$, 
solid line: $\Lambda=10^{-51}m^{-2}$.
Waves with $10^{3}\text{Hz}<w <10^{-10}\text{Hz}$ cannot be practically
plotted in the relevant $Z$-range. Here $w=4\cdot 10^{-16}$Hz}
\end{figure}

\section{Timing residuals in Pulsar Arrays}
Pulsars are very stable clocks with periods ranging from milliseconds up to about 10 seconds.
About 600 pulsars are known. The average period is 0.65 s.
Fast rotating pulsars can be more regular than atomic clocks and disruptions 
of the order of $1 \mu$s can be measured on Earth.

The passage of a gravitational wave disrupts this array of clocks and indeed PTA's may provide the
first {\em direct} evidence of gravitational waves in $< 10$ years. The idea behind the PTA collaborations 
is to detect the correlated disruption of the periods measured 
for a significant number of pulsars due to the passing of a gravitational wave through the system 
\cite{hobbs1,hobbs2,hobbs3,lee}.
PTA are suitable detectors for low frequency GW, i.e. for the range $10^{-9}$ Hz 
to $\le w\le 10^{-7}$ Hz \cite{hobbs1} and the signal is expected to follow a power law \cite{hobbs2,jenet}. 
A key problem in making predictions is modeling in a realistic way the wave functions produced in the different 
sources, in particular the value of the amplitude of the metric perturbation $h$ is a free parameter 
in principle. Some bounds in the range of $10^{-17}\leq h\leq 10^{-15}$ have been set already \cite{jenet}.

Rough estimates from the expressions in the
previous section indicated that corrections  of $O(\sqrt{\Lambda})$ can be relevant for pulsar timing arrays
(typically situated at $L\sim 1$ kpc) being disturbed by extragalactic binary black hole systems (typically
at $l\sim 100$ Mpc or more).

If $\phi_0(t)$ is the field phase of the pulsar, the pulsed emission measured with an Earth-based radio telescope
will be 
\be
\phi(t)= \phi_0(t-\frac{L}{c} -\tau_{E+S}(t)-\tau_{GW}(t))
\ee
where $\tau_{E+S}$ are local corrections due to the movement of the Earth and solar system
and $\tau_{GW}(t)$ is the shift due to the passage of a GW.

Let us consider the observational set-up shown in the figure
\begin{figure}[htbp]
\centering
\includegraphics[scale=0.7]{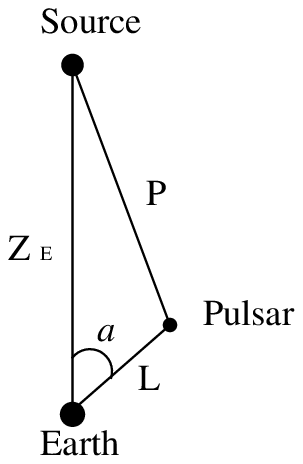}
\end{figure}
The field phase shift due to the GW will be approximately given by\cite{residual}
\be
\tau_{GW}(t)=  - \frac12 \hat n^i \hat n^j H_{ij}(t),
\ee 
where
\be
H_{ij}(t) = L \int_{-1}^0 dx \, h_{ij}(t+Lx, \vec P + L(1+x)\hat n) 
\ee
and $\vec P$ is the pulsar location,   $\hat n= (- \sin\alpha,0,\cos\alpha)$ and  $Z_E= c z \sqrt{\frac{3}{\Lambda}}$
($z$ is the redshift).
Here and in what follows we do not take matter into account, that is $\Omega_M=0$. We shall return to 
this point later.

It is interesting to note that this expression (which includes the leading correction anyway) is valid
only if the time components of the perturbed metric are all zero. This is so at order $\sqrt \Lambda$, which 
by far dominates, but ceases to be true when one considers the ${\cal O}(\Lambda)$
terms indicated in (\ref{waveleading}). See \cite{BEP} for the complete expressions.

To keep things as simple as possible let us assume that a remote merging of galaxies eventually
leads to the merging of their central black holes. Characteristically, the two black holes have
very different masses and therefore we can think of one object orbiting around the most massive one, and
eventually collapsin onto it.
The problem is therefore a keplerian one in essence, with approximate spherical symmetry.
The two spiraling black holes produce GW with a characteristic time
of emission that is of the order of one to several years and the period of the signal ranges from days to months.
The coordinates where the emission is just a collection of a few harmonics will of course
be SdS coordinates. Consider the simplest possible case (just one harmonic):
\be
h^{SdS}_{\mu\nu}=\frac{1}{r}\left(E_{\mu\nu}\cos[{w(t-r)}]+D_{\mu\nu}\sin[{w(t-r)}]\right)
+ O(\Lambda)
\ee
When we say days or months we have to be definite about which clock we are talking about; namely in 
which coordinate system $t$ and $r$ are measured in the previous equation. It should be obvious that they 
refer to the coordinates associated to the supermassive black hole (located 
at $r=0$). It is only in those coordinates that the emission of GW is periodic. (Of course it is not 
{\em exactly} periodic as the smallest black hole loses energy and eventually collapses and the 
emission of the GW cannot be really attributed to the point $r=0$, but this does 
not change the essence of the argument because the uncertainty in the point of emission is much
smaller than other magnitudes relevant for the discussion.) In conclusion, the previous equation is the 
distortion to the metric produced by GW in SdS coordinates.

If the metric would be exactly Schwarzchild, that is without any cosmological constant at all, the metric 
would asymptotically become Minkowski and these coordenates are roughly speaking also the ones of a remote observer
(remember that we have neglected the presence of matter completely). However, this is certainly not the
case if $\Lambda\neq 0$ because then the metric is not asymptotically flat. Then it is clear that a 
terrestrial observer is not using the coordinates $r$ and $t$ but rather describes their observations of the
universe (in particular the distant place where the black hole merging took place) using FRW coordinates 
$R$ and $T$.

As we learned in the previous sections then the GW becomes in FRW coordinates
\be\label{wavefrw}
\bal
&h^{FRW}_{\mu\nu}=\frac{E_{\mu\nu}}{R}\left( 1+\sqrt{\frac{\Lambda}{3}}T\right)
\cos\left[{w(T-R)+w\sqrt{\frac{\Lambda}{3}}\left(\frac{R^2}{2}-TR\right)}\right] \\
& + \frac{D_{\mu\nu}}{R}\left( 1+\sqrt{\frac{\Lambda}{3}}T\right)
\sin\left[{w(T-R)+w\sqrt{\frac{\Lambda}{3}}\left(\frac{R^2}{2}-TR\right)}\right]+ O(\Lambda).
\eal
\ee
As we discussed in detail in the previous section there are corrections of order $\Lambda$ to the 
expression for the waves in SdS coordinates, but after changing to FRW coordinates, these corrections
will still be of order $\Lambda$. It is really the change from SdS to FRW coordinates that matters
and introduces corrections of order $\sqrt\Lambda$.

From the pulsar to the Earth the electromagnetic signal follows the trajectory given by the line of sight 
$\vec R(x)= \vec{P}+L(1+x)\hat{n}$. $L$ is the comoving distance (replacing it by the ruler distance 
makes no significant differences). 
\be
R(x)=\sqrt{Z_{E}^2+2xLZ_{E}\cos{\alpha}+x^2L^2}\simeq Z_{E}+x L\cos{\alpha},
\ee
In the usual treatment, the cosmological constant is neglected and 
the effect of $\Lambda$ would taken into account only through the  redshift 
$w \to w_{eff} =w/(1 + z)$. The important question is of course whether
$\Lambda$ is really relevant after all.

\section{Observing cosmological constant effects in gravitational waves}
Let us define
\be
\bal
&\Theta(x,T_E,L,\alpha,\beta,Z_{E},w,\Lambda)\equiv w(T_E+\frac{L}{c}x-\frac{Z_{E}}{c}-x\frac{L}{c}\cos{\alpha})\\
&+w\sqrt{\frac{\Lambda}{3}}\left(\frac{(\frac{Z_{E}}{c}+
x\frac{L}{c}\cos{\alpha})^2}{2}-\left(T_E+\frac{L}{c}x\right)\left(\frac{Z_{E}}{c}+x\frac{L}{c}\cos{\alpha}\right)\right).
\eal
\ee
Then
\be\label{tresidual}
\bal
H\equiv \tau_{GW}(T_E)=&-\frac{1}{2}\frac{L\varepsilon}{c}\left(\sin^2{\alpha}\cos^2{\beta}+2\sin{\alpha}\sin{\beta}\cos^2{\beta}-\sin^2{\alpha}\sin^2{\beta}\right)\\
\times &\int_{-1}^{0}dx\frac{1}{(Z_{E}+x L\cos{\alpha})}\left( 1+\sqrt{\frac{\Lambda}{3}}(T_E+\frac{L}{c}x)\right)\left(\cos\Theta +\sin\Theta\right),
\eal
\ee
where $\varepsilon$ is a characteristic GW amplitude. All the variables have already been defined
except  $\beta$ that corresponds to the azimuthal angle of the pulsar 
referred to the plane perpendicular to the line Earth-source.

The first indication that the cosmological constant matters comes from considering the 
theoretical dependence of the signal 
on the angle $\alpha$ subtended by the source and the pulsar, the Earth being the vertex.
This is shown in the figure where an enhancement is found at low angles (at least for the values
selected for the astrophysical parameters)
\begin{figure}[htb]
\centering
\includegraphics[scale=0.35]{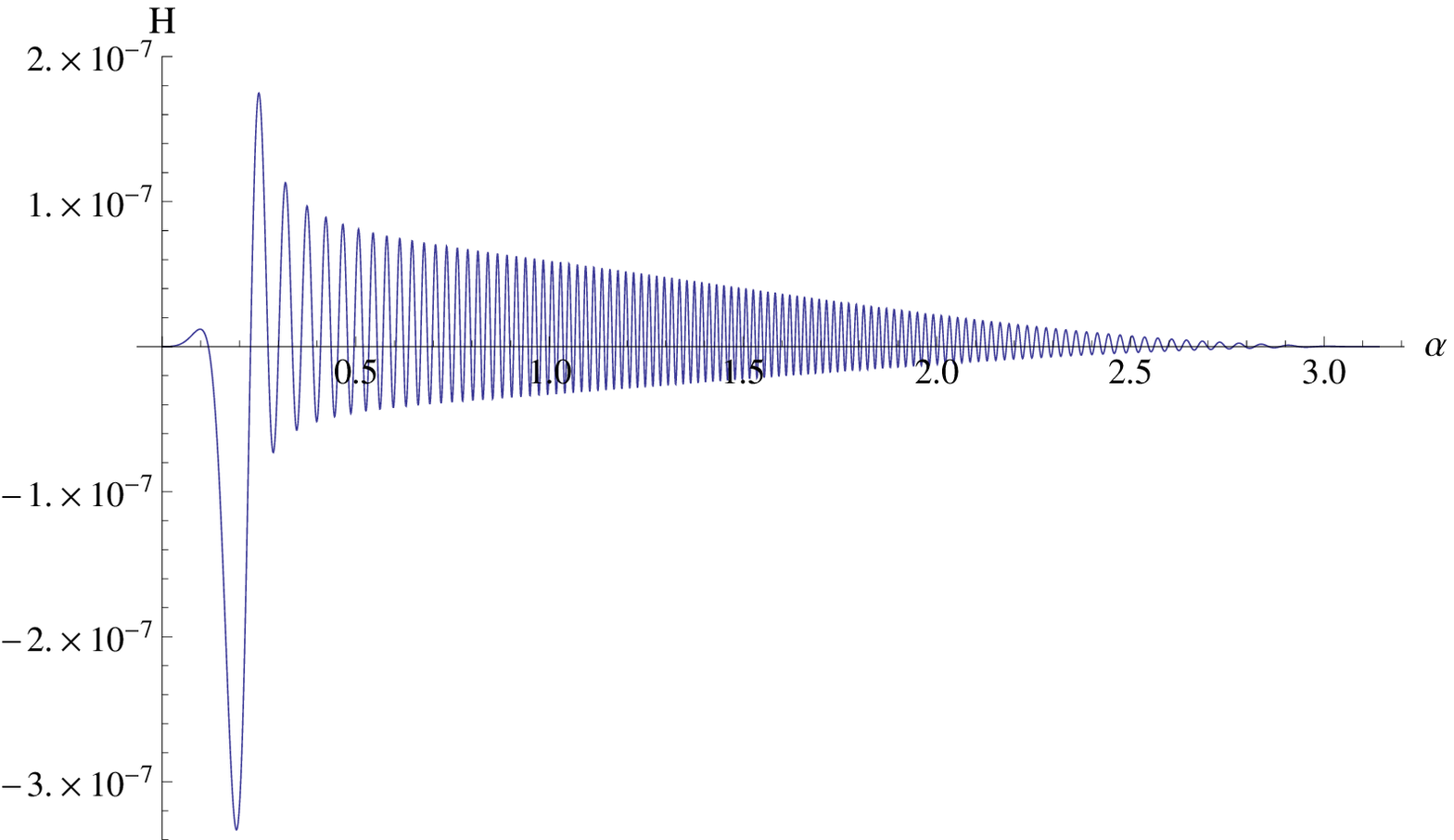}
\includegraphics[scale=0.35]{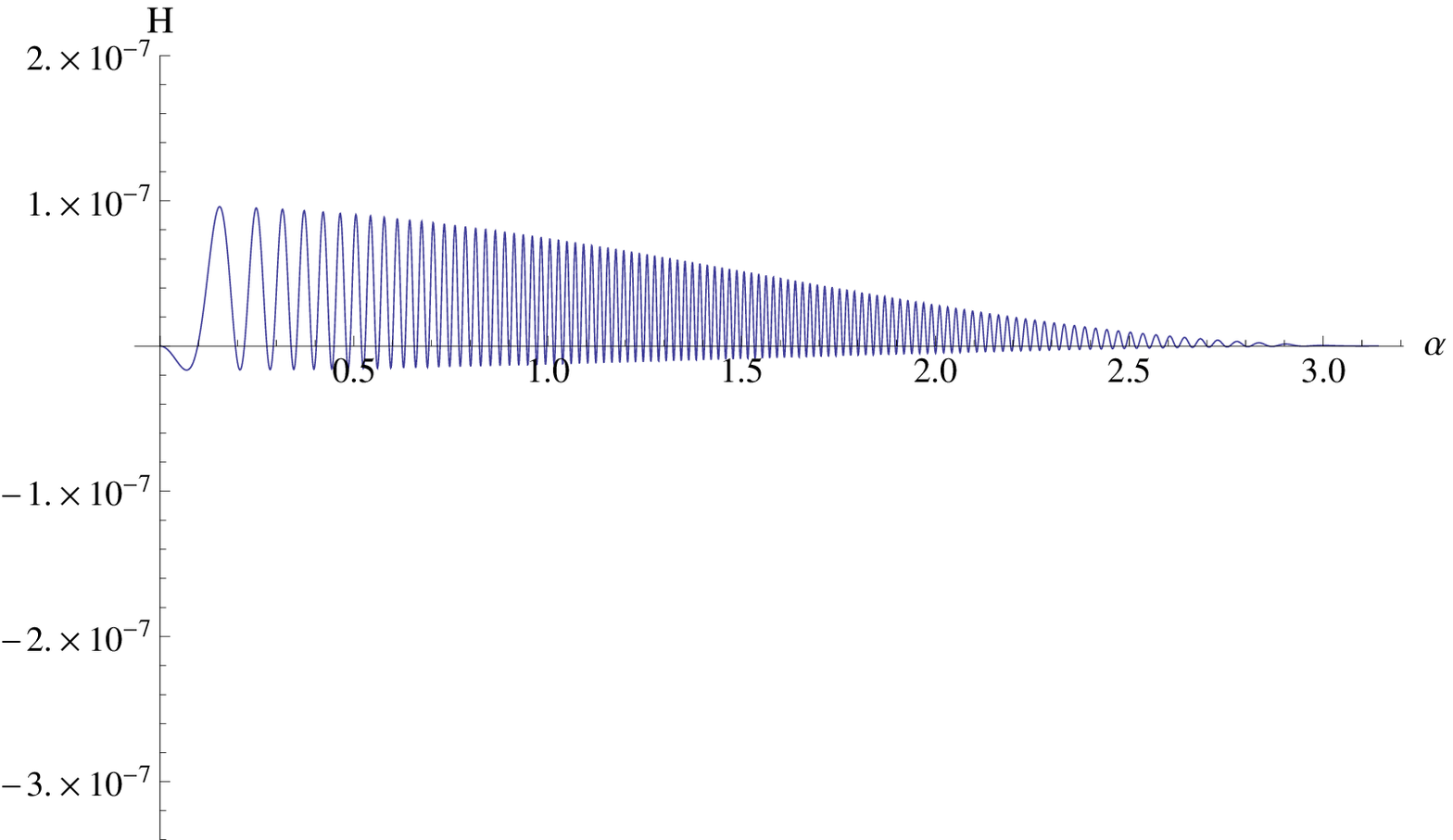}
\caption{On the left the raw timing residual for $\Lambda=10^{-35}s^{-2}$ as a function
of the angle $\alpha$ subtended by the source and the measured pulsar as seen from the observer. 
On the right the same timing 
residual for $\Lambda=0$. In both cases we take $\varepsilon=1.2\times 10^{9}m$, $L=10^{19}m$ 
and $T_{E}=\frac{Z_{E}}{c}s$ for $Z_{E}=3\times 10^{24}m$; with these values $|h|\sim \frac{\varepsilon}{R}\sim 10^{-15}$ 
which is within the expected accuracy of PTA. Similar results are obtained for other close values of $T_{E}$ }
\label{time2}
\end{figure}

Let us now define the following statistical significance
\be
\sigma=\sqrt{\frac{1}{N_{p}N_{t}}\sum_{i=1}^{N_{p}}\sum_{j=1}^{N_{t}}\left(
\frac{H(T_E^{i,j},L_{i},\alpha_{i} ,\beta_{i} ,Z_{E},w,\varepsilon,\Lambda)}{\sigma_{t}}\right)^2}
\ee
where $\sigma_{t}$ is the accuracy with which we are able to measure the pulsar signal period. 
We take $\sigma_{t}\simeq 10^{-6}s$ (this is the average of the 
best measured pulsars included in the International PTA Project \cite{hobbs3}).

We assume an observation time of  3 years, 
starting at the time the signal is $10^{16}s$ old 
(time of arrival at our Galaxy) and observations every 11 days
($N_{t}=101$); $10^{16}s\leq T_{E} \leq 1.00000001\times 10^{16}s$. The coalescence 
times of supermassive black holes is taken to be $O(10^{7})$ s; that is about one year
(much shorter time scale than the time of arrival of the perturbation to the local system).

The galactic latitude and longitude of each pulsar 
are transformed to ($\alpha$,$\beta$), where $\alpha$ is the angular separation 
between the line Earth-GW source and the line Earth-pulsar.

We plot $\sigma(\alpha)$  using a set of 5 fixed 
pulsars supposed to be exactly at the same angular separation from a source the position of which we 
vary. This could be done for any set of five pulsars. The position of the peak does not 
depend on $L_{i}$ and $\beta_{i}$. We use the pulsars 
which are all close to each other at a distance $L\sim 10^{20}m$: 
\begin{table}[htb]
\centering
\begin{tabular}{|c|}
\hline
Pulsars from the ATNF Catalogue\\
\hline
  J0024-7204E	\\
  J0024-7204D	\\
  J0024-7204M	\\
  J0024-7204G	\\
  J0024-7204I	\\
\hline
\end{tabular}
\caption{}
\end{table}

\noindent
The results are summarized in the following figures:
\begin{figure}[htb]
\centering
\includegraphics[scale=0.55]{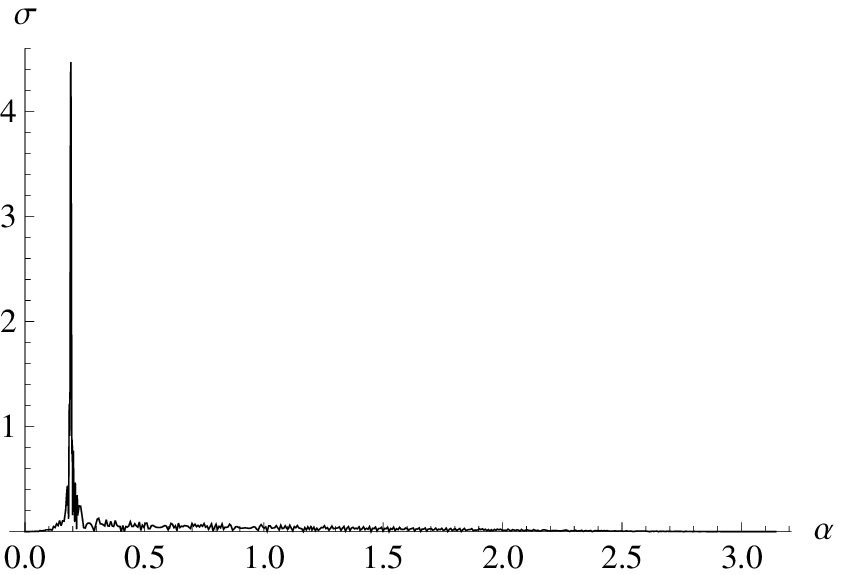}
\includegraphics[scale=0.55]{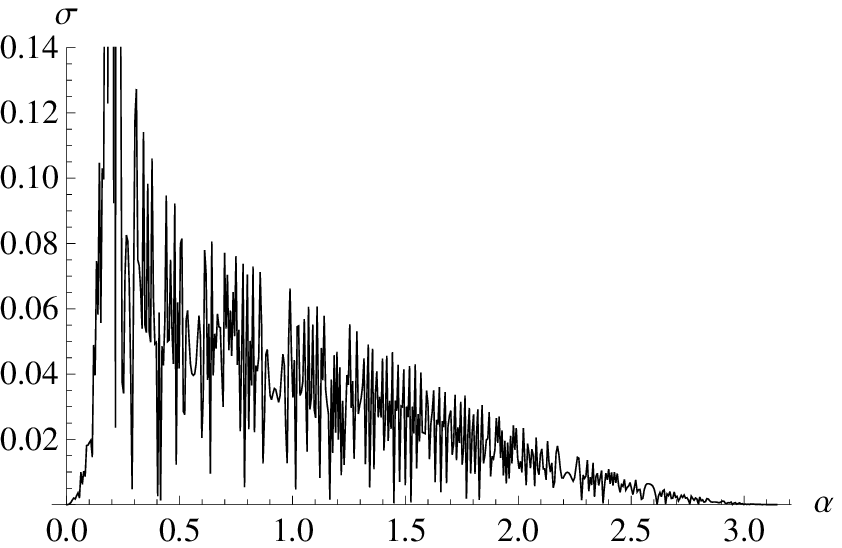}
\includegraphics[scale=0.55]{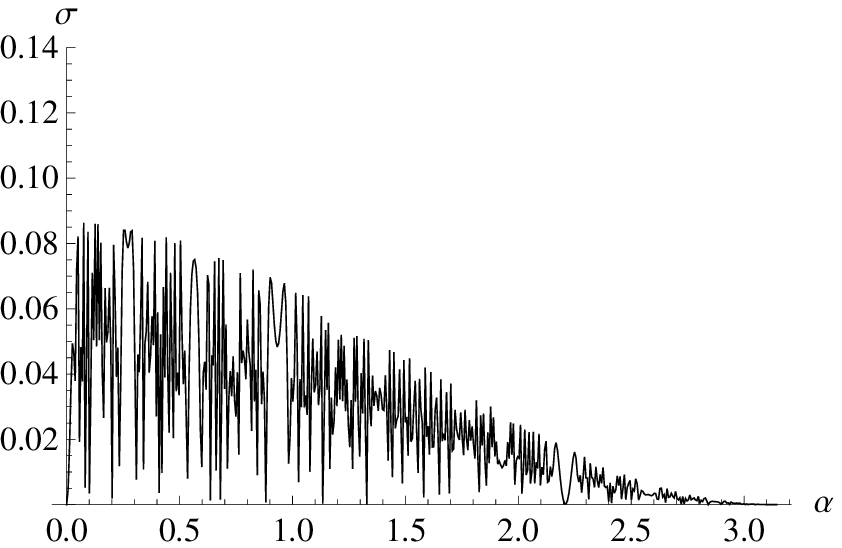}
\caption{$\sigma(\alpha)$ for $\Lambda=10^{-35}s^{-2}$ (left). 
Zoom on the lower values for $\Lambda=10^{-35}s^{-2}$ (middle) with the
prominent spike removed (notice the very different vertical scale, and comparison 
to the $\Lambda=0$ case (right).}
\end{figure}

As is obvious from the figure, the signal shows a remarkable similarity between the cases 
$\Lambda=0$ and $\Lambda\neq 0$ except for the very prominent spike that the case
shown for $\Lambda=10^{-35}s^{-2}$, the currently preferred value for $\Lambda$, at a relatively
low angle. 

Now take a list of observed pulsars well distributed in the galaxy. The angles $(\alpha,\beta)$ are calculated 
for all of them considering two hypothetical sources of GW. One located 
at galactic coordinates $\theta_{S1}=300^{\circ}$, $\phi_{S1}=-35^{\circ}$ and another 
at $\theta_{S2}=4^{\circ}$, $\phi_{S2}=10^{\circ}$. 

We order them from the lowest $\alpha$ to the largest. We 
group them in sets of five pulsars. We consider 27 sets of 5 pulsars; that is a list of 135 pulsars. 
For each set we calculate the significance
\be
\sigma_{k}=\sqrt{\frac{1}{5\cdot 101}\sum_{i=1}^{5_{k}}\sum_{j=1}^{101}\left(
\frac{H(T_E^{i,j},L_{i},\alpha_{i} ,\beta_{i},10^{24},10^{-8},1.2\times 10^{9},10^{-35})}{10^{-7}}\right)^2}
\ee
and plot it as a function of the average angle of the set, $\bar{\alpha}_{k}=\sum_{i=1}^{5_{k}}\frac{\alpha_{i}}{5}$ 
with $1\leq k \leq 27$. 
\begin{figure}
\centering
\includegraphics[scale=0.35]{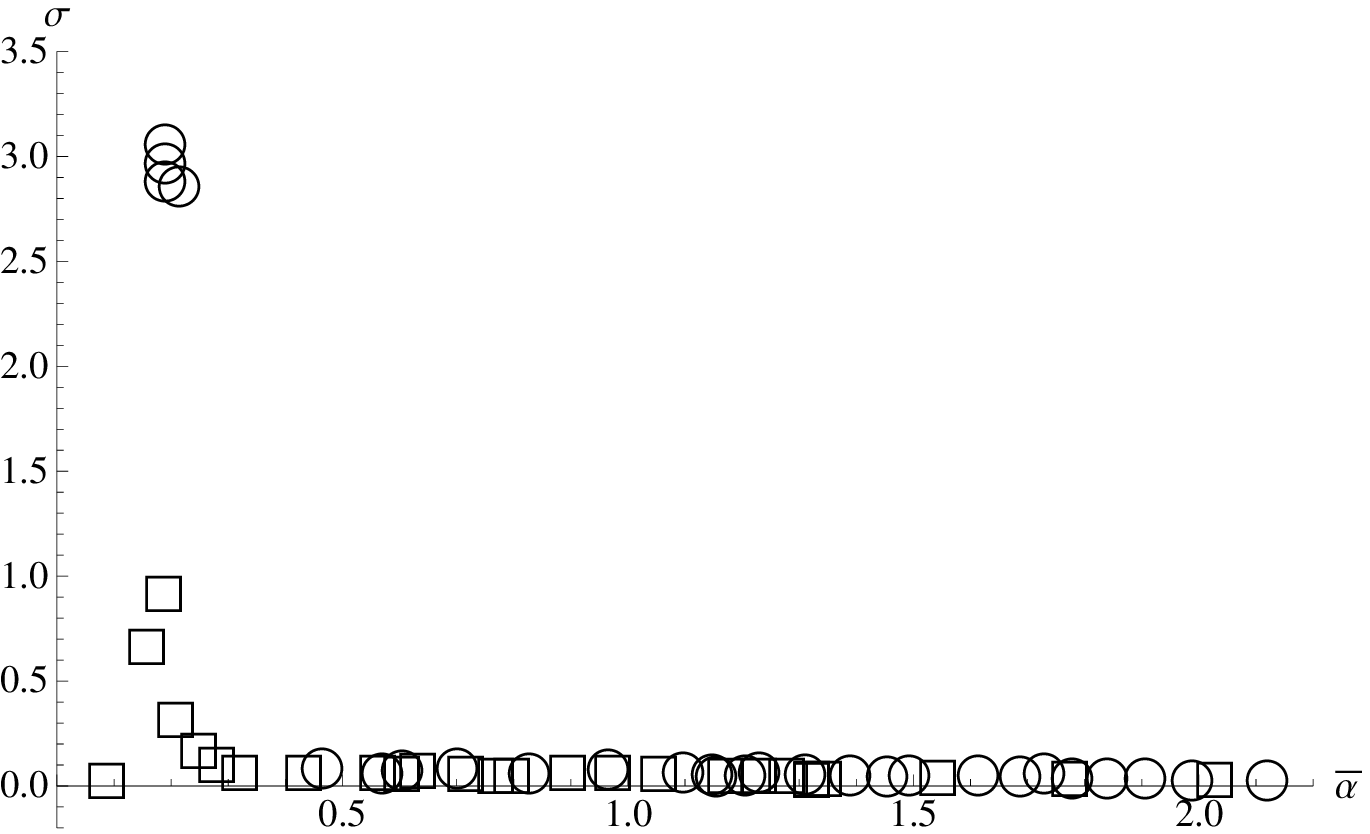}
\includegraphics[scale=0.35]{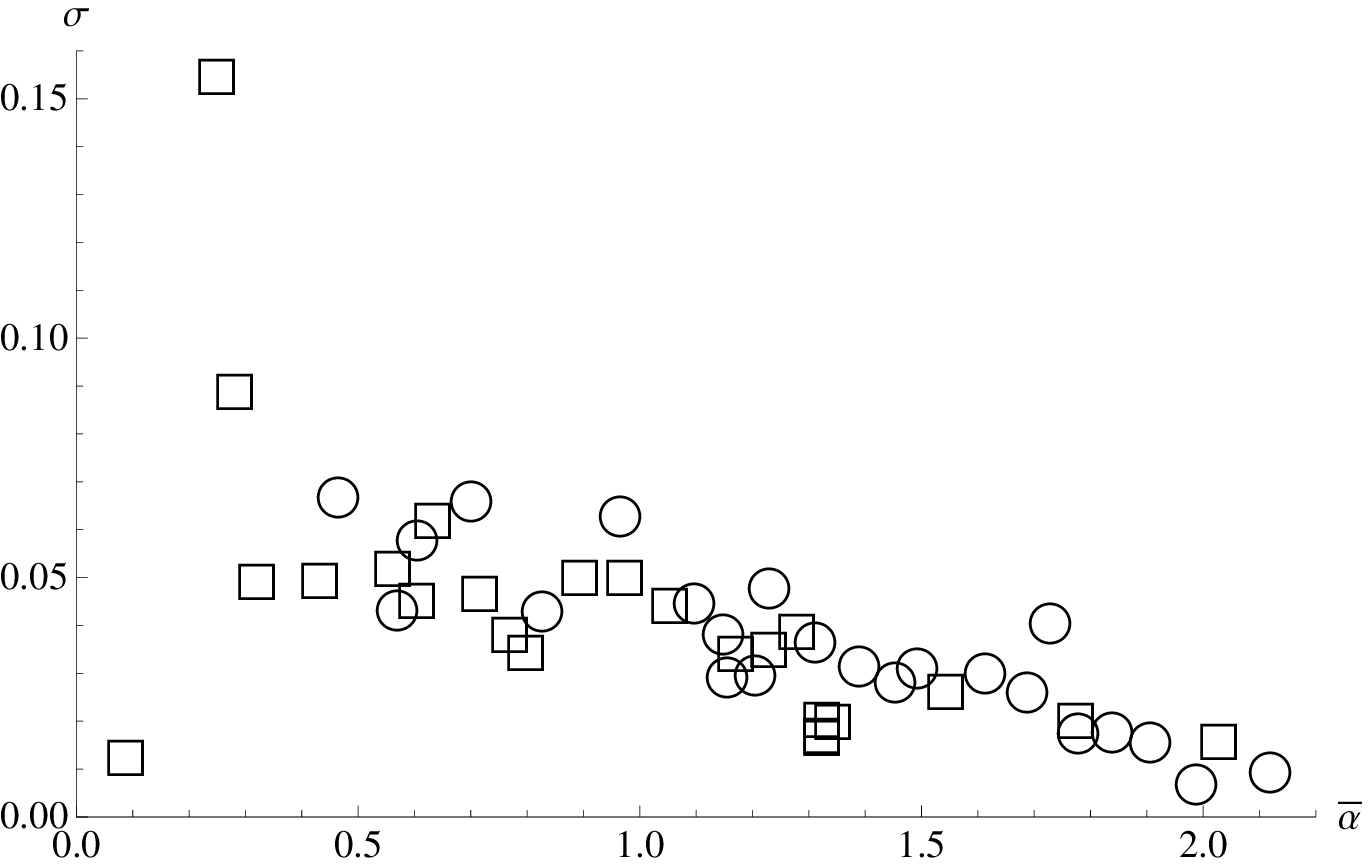}
\includegraphics[scale=0.35]{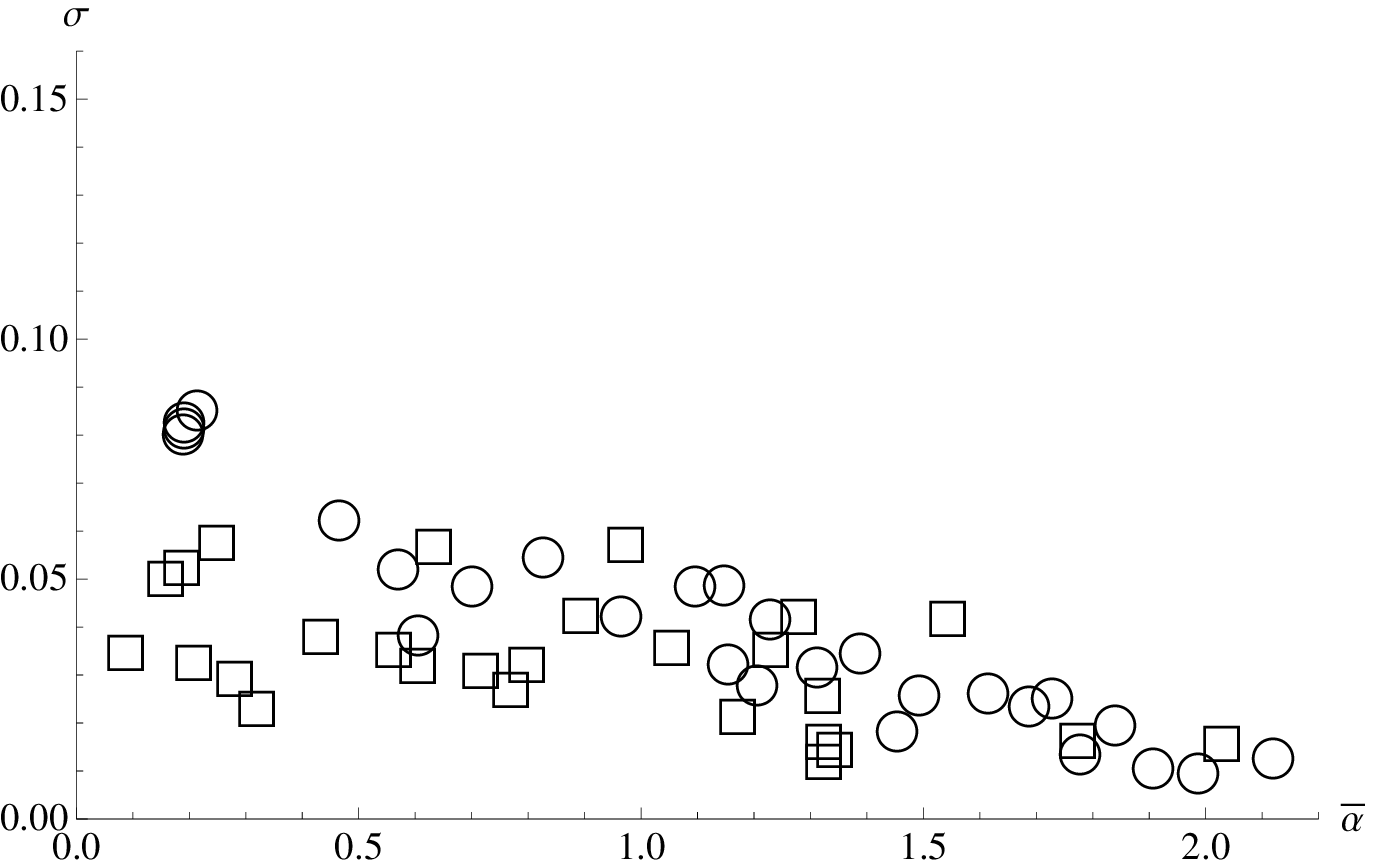}
\caption{Plot of $\sigma_{k}(\bar \alpha_{k})$, $k=1,27$. $\Lambda=10^{-35}s^{-2}$. Circles correspond to Source 
1 and squares to Source 2. Full range is showed on the left, zoom on the lower values for $\Lambda=10^{-35}s^{-2}$ 
is shown in the middle (again the spike is removed -vertical scale is different) and 
comparison to $\Lambda=0$ is show on the right figure, respectively.}
\label{picp}
\end{figure} 
We think that the figures speak by themselves. It is clear that PTA observations that aim at observing
the `normal' GW spectrum should absolutely see the `abnormal' enhancement at a given value of the
angle.

\section{Why this enhancement?}
In order to understand why this effect comes about
let us examine the behaviour of the differential timing residual as we move along the line of sight.
\begin{figure}[htb]
\centering
\includegraphics[scale=0.42]{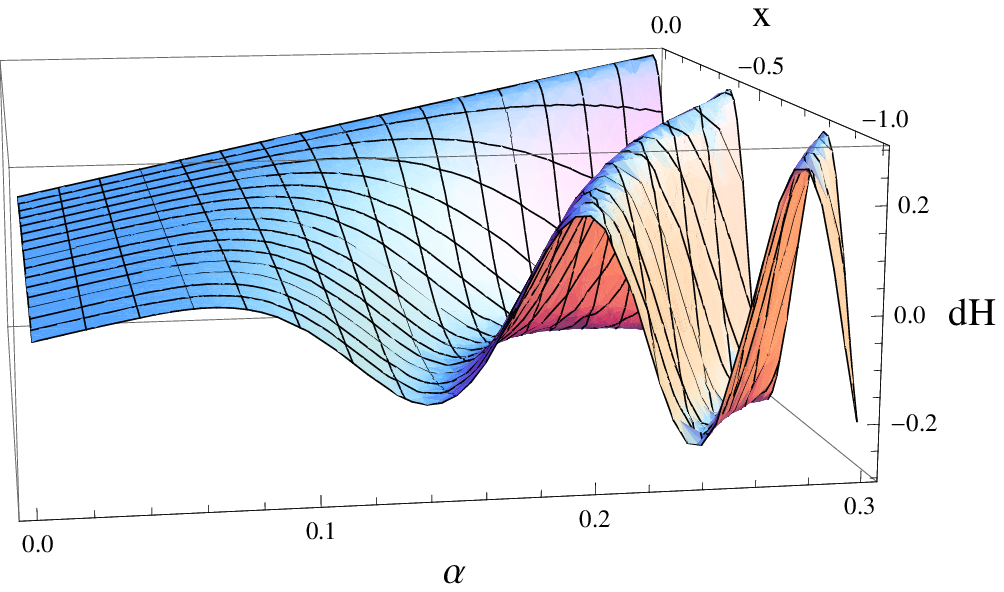}
\includegraphics[scale=0.42]{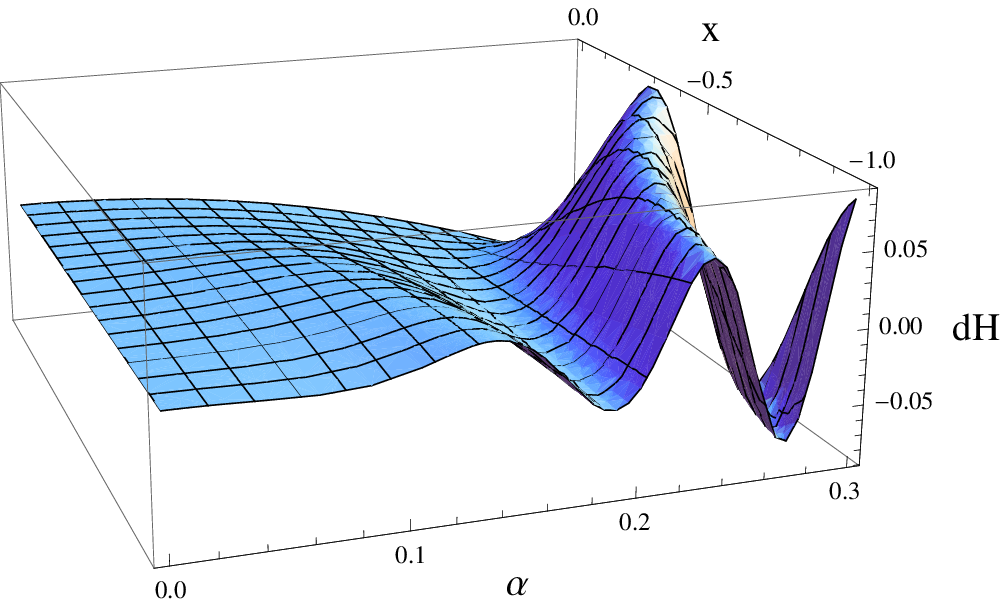}
\includegraphics[scale=0.42]{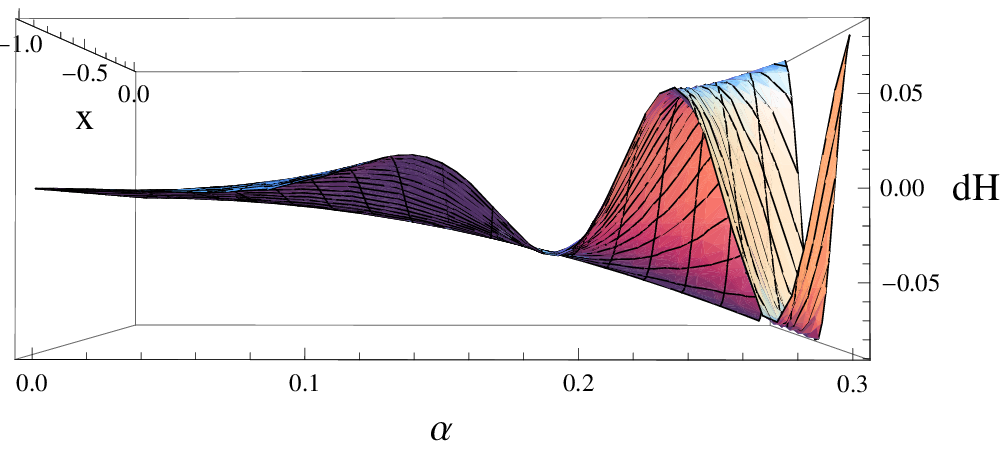}
\caption{Evolution of the phase of the GW along the line of sight to the pulsar for different angles.
Left figure  $\Lambda=0$. Middle:  $\Lambda=10^{-35}s^{-2}$. Right: same as the previous one but rotated so
as to show the `valley' of stationary phase, which is absent without the cosmological constant.}
\end{figure}
The window in the angular variable corresponds to a `valley' where the phase is (nearly) stationary. At this
value of the angular variable $\alpha$ is where the enhancement takes place.

Among all the dependencies, and when the distance to the source is well known, the most relevant appears to
 be the one related to the value $\Lambda$. The position of the peak depends strongly on the value of $\Lambda$. 
It moves towards the central values of the angle for larger values of the cosmological constant. 

The position of the `valley' (and therefore $\Lambda$) can be found analytically in two ways
\begin{enumerate}
\item
Looking for the stationary phase condition
\item
Examining the behaviour of the Fresnel functions and prefactors obtained after integration
\end{enumerate}
The prefactor becomes quite large for a specific value of the parameters involved. This particular 
value renders the Fresnel function close to zero and the product is
a number close to 2. Away from this point the net result is small. 

Using the series expansion of the Fresnel functions at first order we are able to obtain an 
approximate analytical expression for the relation $\Lambda(\alpha)$
\be\label{lambalpha}
\Lambda(\alpha)=\frac{12 c^2 \sin ^4\left(\frac{\alpha }{2}\right)}{\left((c T_{E}-Z_{E})\cos \alpha  +
Z_{E}\right)^2} \simeq \frac{12 c^2 \sin ^4\left(\frac{\alpha }{2}\right)}{Z_E^2},
\ee
\begin{figure}[htb]
\centering
\includegraphics[scale=0.5]{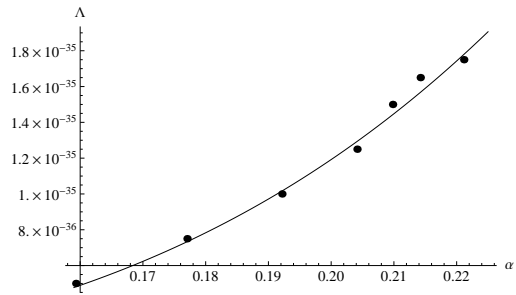}
\caption{$\Lambda(\alpha)$ obtained numerically from the positions of the peaks in the $\sigma(\alpha)$ plots for 
different values of the cosmological constant (dots) and obtained analytically from an approximation of 
the Fresnel functions involved in the timing residual (line).}
\end{figure}

However it is possibly more enlightening to understand the effect via the first explanation (which of course
is mathematically equivalent to the second, but clearer from a physical viewpoint). For the sake of this
discussion consider a scalar wave with the same arguments as (\ref{wavefrw}), parametrized along 
a null geodesic using (we set here again $c=1$ and choose the initial phase to be convenient for our discussion)
\be
t = T + xL, \quad r = R + xL \cos\alpha, \quad -1 < x < 0, 
\ee
i.e.
\be
\phi= \frac{1}{R + xL \cos\alpha}\cos\left[
wxL(1-\cos\alpha) + w\sqrt{\frac{\Lambda}{3}}\left(\frac12(R + xL \cos\alpha)^2- 
(R + xL \cos\alpha)(T + xL)\right)\right].
\ee
If we take the derivative of the argument with respect to x and keep only the terms that really
matter for the relevant values of the parameters, we get
\be
\frac{d\phi}{dx}\simeq w \left(L (1- \cos\alpha) + \sqrt{\frac{\Lambda}{3}}
(-LT\cos\alpha-LR(1-\cos\alpha))\right).
\ee
As just stated, in the derivative we have neglected some terms proportional to $x$ that are very small. 
Now using that $T\simeq R$ to a very good precision, it can be shown this expression 
has a zero at the value of $\alpha$ corresponding to
\be
\sin^4\alpha\simeq \frac43 R^2 \Lambda
\ee
that for small values of $\alpha$ agrees with (\ref{lambalpha}). 

The enhancement's angular position corresponds very approximately to a stationary line of the wave. 
The value of the phase of the wave remains practically constant (with a wave amplitude different from
zero) along the straight line from the pulsar to the Earth for that particular value of $\alpha$. So
the integrated timing residual is maximized for this angle.

The limit $\Lambda=0$ also has such stationary phase, it obviously corresponds to $\alpha=0$ where 
the light signal from the pulsar and the GW travel in phase towards us, but the relevant point is
that the angle $\alpha$ where the enhancement is observed gives a very direct determination
of the cosmological constant, a remarkable property that was not noticed before.
As we see the only unknowns are the distance $R$ and 
$\Lambda$ itself, which is remarkable. In particular the effect is independent of the frequency, and
also of $L$, the distance to the given pulsar used as a clock.
Of course if we trust
that $\Lambda$ is a universal constant value this expression can be used to obtain $R$, the distance
from the Earth to the source of GW.

\section{A critical look}
Although the results presented are quite interesting there are several points on which they 
are manifestly incomplete and therefore the conclusions one may draw can be criticized.
\begin{enumerate}
\item
The wave front used is not realistic. Indeed, we have assumed a periodic wave with only one harmonic. However, 
it would be quite straightforward to extend the analysis to the different Fourier components of a given signal and
the essence of the discussion would be exactly the same. Because the effect is relatively 
insensitive to the frequency\cite{EP}
and depends mostly on $Z_E$  and $\Lambda$ we would not expect substantial changes 
in the conclusions. This is surely
the least relevant of the possible objections.

\item
The time dependence of the signal will average out the effect. We have sometimes been asked why the effect is not
averaged out. This objection seems to
arise from a misunderstanding of formula (\ref{tresidual}). It is true that the period of observation
is typically much longer than the typical period of the black hole spiraling. However as is obvious from the explanation
provided in the previous section, the frequency itself plays no role in the construction of the `valley' corresponding
to the stationary phase condition. The only time that plays a role at all is $T_E$, and indeed the signal would average
out after a time  $>> T_E$, but this is a very long time, many orders of magnitude longer than the
observational period.  

\item
The matter distribution in the galaxy will perturb the signal and finally wash it out. This is a valid concern as
local distortions of the background metric certainly influence to some extent the propagation of GW. Indeed although
globally $\Lambda$ contributes more to the matter-energy budget of the universe, in the Galaxy the balance is obviously
reversed (in fact it will be dominated by the dark matter component of the halo). However, this and similar 
effects arise from non-linear terms in the equation governing GW propagation. In fact they are similar 
in effects, but locally
larger, than the term $\Lambda h_{\mu\nu}^W$ that was neglected in the discussion. But of course they are relevant only
in the vicinity of the Galaxy. When the GW arrive at the Galaxy, anharmonicity is already well developed and
changes with respect to the $\Lambda=0$ case noticeable. Although the fine details remain to be worked out, it seems
extremely unlikely that these effects could be really significant.  

There are two more effects that would need to be taken into account due to the matter distribution in the Galaxy.
First of all, the time for the pulsar(s) signal to reach the Earth in the absence of the GW is not exactly $L/c$ as
there are local perturbations on light propagation. However, these are very slowly varying with time ---except
for the peculiar motion of the Earth and Sun, separately taken into account. This is therefore in any case an  
additional term that does not really change the effect.

A last ingredient to be taken into account is to consider the modifications brounght about by the matter 
distribution in the Galaxy in the precise relation between the cosmological time and the observer's time defined
by a clock on the Earth. It does not seem that this can have any effect at all in the kind of signal predicted.

\item
Background noise will make the signal undetectable. In the universe there is not a single source of 
gravitational waves of this kind and we should expect that the PTA detect many such events \cite{noise}.
Unfolding the signal is
of course non-trivial but it does not seem to be worse in our proposal than in the usual case. In practice 
this is a serious difficulty in all GW detectors, not just PTA. In fact the predicted enhancement at a given angle
should facilitate things.

\item
The matter density has been neglected. This is by far the most serious shortcoming of our proposal. So far we
have not taken into account the fact that $\Omega_M$ in non-zero in the universe and we have not taken into account
the effects due to the masses of the coalescent black holes. Both effects should definitely be taken into account.

\end{enumerate}

\section{Conclusion and outlook}
In this presentation we have summarized our recent joint work with J. Bernabeu and D. Puigdom\`enech on the issue of
GW in de Sitter space-time. We study a very definite problem: how a GW produced in the merging of two black holes
propagates in de Sitter space-time and how is detected. Conventionally, the effect of the cosmological constant
is included via the usual redshift in the wave frequency. We criticize this simple approach and see that there
is a lot more to it.

In order to understand the role of $\Lambda$ in the propagation of GW using a 
linearized approximation seems essential. This linearization procedure is of course the common one to treat small 
perturbations around Minkowski space-time (in a sense, this is necessary in order to {\em define} a gravitational wave
and also to keep track of the $\mathcal{O}(\Lambda)$ corrections.  
Another line of work studies the small perturbations around a FRW metric corresponding to primordial GW. It should be clear
to the reader that this is not what interests us here. Here the problem to study is
completely different and in order to keep track the different orders in $\Lambda$ it is useful
to linearize {\em both} the background and the GW perturbation. 

To understand the propagation of GW the different coordinate systems involved have to be thoroughly understood. As
a result of the anlaysis GW that are harmonic when produced in the merging of two supermassive black holes
turn out to be anharmonic in cosmological coordinates, the ones where we observe.

In fact the modifications appear to be very relevant for PTA. There is a dramatic enhancement for a given 
value of the angle $\alpha$ subtended by the source of the GW and the corresponding pulsar and that 
depends strongly on the value of $\Lambda$.

This work is still very preliminary, for instance the fact that $\Omega_M\neq 0$ has not been considered yet. But
the possibility that GW can be used in a not too distant future to perform a `local' measurement of $\Lambda$ seems
fascinanting and promising.

\section*{Acknowledgements}
It is a pleasure to thank my collaborators J. Bernabeu and D. Puigdom\`enech. The financial support from grants 
FPA2010-20807 (MICINN),  2009SGR502 and Consolider grant CSD2007-00042 (CPAN) is gratefully acknowledged.  
I would like to thank A. Andrianov, V. Andrianov, S. Afonin and all the other colleagues at the University of 
Saint Petersburg for organizing the II Russian-Spanish joint workshop on Particle Physics where this work was 
presented and for making it  very enjoyable and fruitful. The author would also like to thank the Faculty 
of Physics of the Universidad Cat\'olica de Chile where these proceedings were written for their warm hospitality.

\bibliographystyle{aipproc}   

\end{document}